# The Virtual Institute for Integrative Biology (VIIB)

Gustavo Rivera[1], Fernando González-Nilo[2], Tomás Perez-Acle[3], Raúl Isea[4]

and David S. Holmes[1]

[1]*Center for Bioinformatics and Genome Biology,*
*Life Science Foundation and Andrés Bello University*
*Santiago, Chile. Gustavo.rivera@bionova.cl, dsholmes2000@yahoo.com*
[2] *Center for Bioinformatics and Molecular Simulation,*
*University of Talca, Talca, Chile. dgonzalez@utalca.cl*
[3]*Center for Bioinformatics, Universidad Católica, Santiago, Chile. tomas@cgb.cl*
[4]*Fundación IDEA. Caracas, Venezuela. 1risea@yahoo.com*

## Abstract

*The Virtual Institute for Integrative Biology (VIIB) is a Latin American initiative for achieving global collaborative e-Science in the areas of bioinformatics, genome biology, systems biology, metagenomics, medical applications and nanobiotechnolgy. The scientific agenda of VIIB includes: construction of databases for comparative genomics, the AlterORF database for alternate open reading frames discovery in genomes, bioinformatics services and protein simulations for biotechnological and medical applications. Human resource development has been promoted through co-sponsored students and shared teaching and seminars via video conferencing. E-Science challenges include: interoperability and connectivity concerns, high performance computing limitations, and the development of customized computational frameworks and flexible workflows to efficiently exploit shared resources without causing impediments to the user. Outreach programs include training workshops and classes for high school teachers and students and the new Adopt-a-Gene initiative. The VIIB has proved an effective way for small teams to transcend the critical mass problem, to overcome geographic limitations, to harness the power of large scale, collaborative science and improve the visibility of Latin American science It may provide a useful paradigm for developing further e-Science initiatives in Latin America and other emerging regions.*

## 1. Introduction

Biology has emerged as one of the major areas of focus of scientific research worldwide. The amount of biological data available is expanding at breathtaking speed. Genbank, one of the largest public repository of DNA and protein sequence information, continues to double in size every 12-15 months and now harbours over 145 billion nucleotides of genetic information for over 240,000 named organisms [1]. DNA sequencing technology continues to improve, pushing costs for sequencing ever lower. The result is that over 600 genomes are now publicly available for analysis and about 2,500 new genome projects are underway as of mid-year 2007 [2]. The interpretation of this primary sequence information coupled with protein structure modelling and high throughput experimentation such as DNA microarray analysis has given rise to a data overload requiring scientists to increase the scale and sophistication in the information technology used for their research. To handle this flood of data nearly 1000 biology databases [3] and about 1200 specialized webservers [4] have been created. However, much of the data is stored in conflicting formats, often with confusing ontologies and is sometimes of dubious quality providing new challenges for the sharing of huge databases in an e-Science environment.

In computational biology, new tools and algorithms have been developed for scientific research that harness the power of the web for accessibility, data sharing and exchange of ideas. The pace of progress is accelerating and scientists increasingly need to rely on trusted directories, portals and useful search engines to keep up with the latest science. Whereas major efforts have been mounted in the USA, Europe and Asia to meet these challenges, less appears to have been accomplished in Latin America and the Virtual Institute for Integrative Biology (VIIB) was developed to exploit this need. The four founding cornerstones of the VIIB are: (i) the Center for Bioinformatics and Genome Biology (CBGB: www.cienciavida.cl), Life Science Foundation in Santiago, Chile, (ii) the Center for Bioinformatics (CBUC: www.cgb.cl), Catholic University in Santiago, Chile, (iii) the Center for Bioinformatics and Molecular Simulation (CBMS: cbsm.utalca.cl), University of Talca, in Talca, Chile and (iv) the Fundación (IDEA: www.idea.gob.ve) in Caracas, Venezuela.

One of the major objectives of the VIIB is to generate e-Science initiatives with a particular focus on biotechnology agendas that have significant impact in Latin America. In addition, there are social and scientific issues in the region that provide special challenges that must be overcome in order to instigate transnational e-Science initiatives. Consequently, strong links have been established with other groups in Latin America. However, recognizing the global nature of e-Science, the VIIB has also mounted collaborations with European and North America groups and is now seeking additional partners to expand into other parts of the world. This paper will outline some of the scientific and human resource development topics being addressed by the VIIB and will discuss some of the e-Science issues that have proved challenging in the Latin American context. Finally, the paper will describe future plans and projections.

## 2. Scientific Agenda

Efforts in the VIIB are being directed principally in bioinformatics, genomics and protein simulation within areas of biotechnology that impact the economy and health of the Southern Cone Latin American region. These include projects related to fruit and wine exports, the salmon industry, forestry and the biorecovery of copper. However, projects are also underway that tackle issues of global significance such as web resource development for mobile gene (insertion sequences) analysis and a web based resource (AlterORF) for improving genome annotation and for aiding in new gene discovery. Three of these efforts will be described herein to serve as models for the type scientific agenda being addressed by the VIIB.

### 2.1. Bioinformatic Support for the Minerals Industry

The production and exportation of metals is a major economic driving force within the Latin American region. Copper accounts for about 40% of the exports of Chile and copper, gold and other metals are also important for the economy of Peru, Argentina and Brazil. The use of microorganisms for biologically-assisted metal solubilization – a process termed biomining or bioleaching - is emerging as a biotechnological alternative to traditional strategies such as smelting (pyrometallurgy) for metal recovery. There are a number of reasons for this: first, as reserves of high grade metal ores continue to decline and standards for environmental protection become more stringent, it becomes increasingly less attractive economically to apply costly and contaminating traditional pyrometallurgy methods. Second, substantial amounts of low grade ores are stockpiled each year as dump material from mining operations and these remain as waste in the absence of an appropriate re-processing technology. Third, pyrometallurgy methods cannot be applied economically to exploit small or remote deposits. In contrast, biotechnology offers an economic and environmentally friendly solution to these problems; it makes possible the recovery of metals at the mine site reducing transport costs, it uses simple infrastructure with low capital and operating costs and it offers better opportunities for environmental compliance, given that, it requires lower energy inputs and is less contaminating (e.g. zero production of noxious gases).

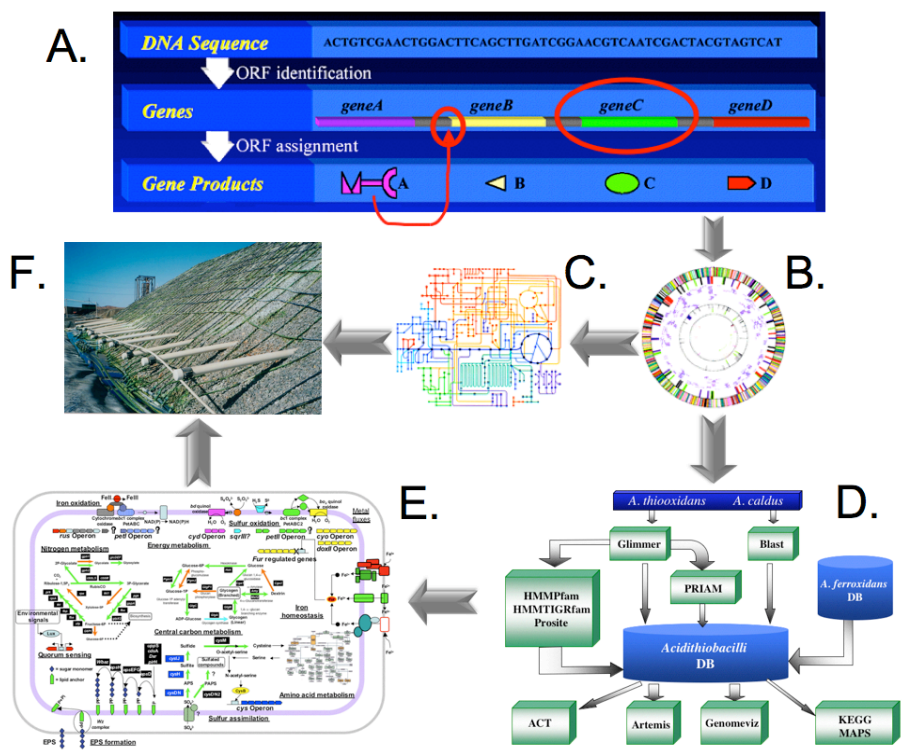

**Figure 1. Workflow for developing databses for biomining applications.** (A) Genomes are annotated, identifying genes and regulatory features. (B) Whole genome maps are prepared. (C) Metabolic models are constructed. (D) Comparative genomics is carried out. (E) Multi-organism comparative metabolic pathways are deduced. (F) Biotechnological application are proposed.

Given the importance of developing biomining in the region, one of the initial objectives chosen for the VIIB was to assist the mining community to leverage biotechnological innovations by providing curated databases of relevant biological knowledge. The genomes of several microorganisms known to be involved in biomining have been sequenced and it is imperative to be able to exploit this information. However, mining companies typically have neither the cyber-infrastructure nor the technical know-how to do this, especially as the relevant genome annotations, metabolic reconstructions and comparative genomics analyses are computationally demanding and require transdisciplinary teams for effective data mining and interpretation. The VIIB provides the community with an overarching database in which genomes have been annotated and preliminary metabolic models have been constructed that forms the basis of a comprehensive framework that not only connects vast amounts of data, but also capture usable knowledge in the form of biologically valid relations that research scientists and engineers can apply (Figure 1).

## 2.2 Depuration of Genome Databases and the Search for Novel Genes

The avalanche of genome sequence data means that automatic tools are required for gene recognition and function prediction. The best annotations of genomes use several automatic gene finding programs simultaneously such as Glimmer and Critica, coupled with statistical evaluations of potential genes using codon usage, dinucleotide frequencies and G+C composition etc. These automatic gene predictions are subsequently followed by extensive manual curation. In less extensively curated genomes, it is thought that a minimum of 10% of genes assigned a putative function might be erroneously identified and even in the best efforts errors of annotation creep in. Because these errors enter public repositories such as Genbank there is serious risk of error propagation and circularity of protein function identification.

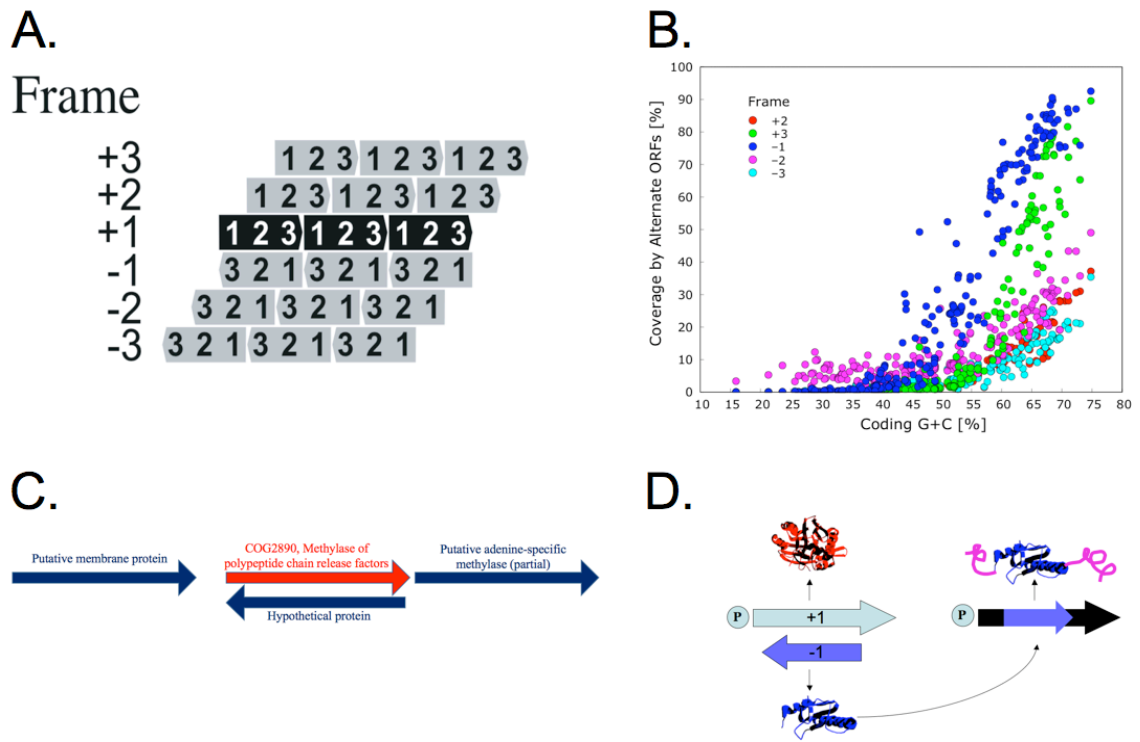

**Figure 2. Information pertaining to the AlterORF Database.** (A) The six possible reading frames of a gene are indicated where frame +1 is defined as that corresponding to the gene sequence deposited in a database such as GenBank. Each group of three numbers corresponds to a codon that specifies an amino acid. There are three reading frames in the forward DNA strand labelled +1 to +3 and three in the reverse complimentary strand labelled -1 to -3. (B) Plot of the frequency of alternate open reading frames (ORFs) as a function of G+C content of the genome. Each point corresponds to the average value of all the genes of one genome. (C) An example where a gene was annotated in GenBank as "hypothetical" whereas it is more likely that it encodes a protein containing a match to the COG database (red arrow). The red arrow corresponds to an alternate ORF discovered by AlterORF. (D) Diagram suggesting how an alternate ORF (blue arrow) can be "captured by a gene (black arrow), for example by transposition, giving rise to a novel hybrid protein that is part original gene and part the inserted alternate ORF.

One of the problems faced by automated genome annotation algorithms is the identification of the correct open reading frame (ORF). Alternate ORFs can arise within genes because, in principle, they can be read by the cellular translational machinary in any one of 6 frames ORF, 3 in the forward direction on one DNA strand and 3 in the reverse direction on the complemementary strand (Figure 2A). Usually, five of these potential reading frames are interrupted by stop codons and so can be eliminated as gene candidates and, even when large alternate ORFs are found, automated gene finding programs can usually identify the correct ORF. However, the frequency of large alternate ORFs is surprisingly high, especially in high G+C genomes, and even the best automated programs make errors in ORF identification (Figure 2B) [5].

The VIIB hosts an open source multi-genome database, termed AlterORF, where pre-calculated alternate ORFs are available in a web interfase for over 600 genomes [6]. Users may also use web based search engines to examine genomes not pre-calculated by AlterORF or they may download tools for in-house analysis. AlterORF identifies all alternate ORFs that could potentially encode known protein domains and motifs linked out to Genbank, Pfam and SwissProt etc. If potential function is predicted in an alternate ORF it suggests that the annotated ORF identified in the original genome and deposited in GenBank could be erroneous (see example in Figure 2C). AlterOrf is proving to be a powerful tool for improving genome annotations by culling alternate ORFs masquerading as real genes. It has been successfully been applied to a suite of genomes involved in human disease.

Interestingly, a few examples have been described of experimentally validated genes that encode protein products from more than one ORF [5] and AlterORF is a rich source for searching for such candidate bi-functional genes. The observation that novel antisense proteins arise in several disease states including cancer and some may provoke auto-immune reactions has prompted further investigation into their physiological role in normal and disease paradigms [reviewed in 7]. AlterORF can also be used to identify potential cases where genes could have evolved by capturing alternate ORFs (Figure 2D).

The creation of the AlterORF database was computationally demanding requiring an all-against-all Blast of the six potential ORFs for every gene of the more than 600 microbial genomes available. Over ten million Blast operations were carried out and the results checked against various databases of protein information. Comparisons involving the use of very demanding algorithms based on Hidden Markov Models. In all, about a terabyte of information has been stockpiled in the AlterORF database.

**2.3 Protein Modelling and Simulation**

The VIIB is hosting a project that investigates how proteins and membrane transport complexes function in extremely acidic conditions. This research could have important biotechnological implications for understanding how microorganisms can extract energy from rocks and how they solubilize copper at pH 1. It could also serve as a platform for the discovery of novel acid stable enzymes for industrial applications, in a similar way that was so successful for the discovery and characterization of thermo-stable enzymes. In addition, the project could reveal fundamental insights into how proteins fold and make protein-protein contact at very low pH. Extensive protonation of residues at pH1 might have required the evolution of novel mechanisms to ensure structural and functional stability and faithful interfacial contact. The project also addresses how membrane transport complexes work when confronted by proton concentrations several orders of magnitude more than in typical neutral pH neutral environments. For example, how do aquoporins that import water into the cell discriminate between water and protons and how do complexes that use protons as an energy source to drive the import or export of nutrients work when confronted by a high proton motive force. A workflow diagram for these projects is shown in Figure 3. The project exhibits the consilience typical of a well structured e-Science initiative, involving a transdisciplinary team of bioinformatics experts, protein chemists and IT specialists located in different geographic areas.

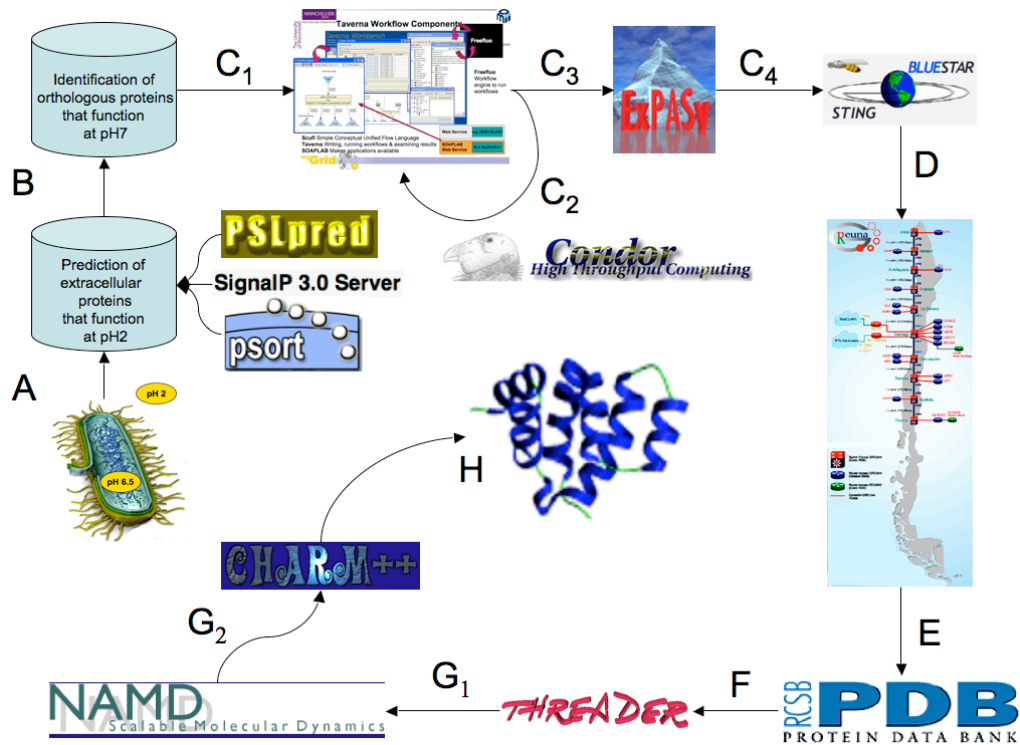

**Figure 3. Work flow for acid stable protein structure and function prediction.** (A) Bioinformatics programs such as pSORT, SignalP and PSLpred are used to predict extracellular proteins from organisms that thrive at pH2 or below. (B) Comparative protein programs recover all similar (orthologous) proteins from organisms that live at neutral pH. ($C_{1-4}$) Proteins to be compared are analyzed via a suite of programs in a Taberna workflow environment using Condor middleware to predict their properties and locate key elements. (D) Resulting information is then sent to collaborators via the REUNA Chilean wideband Internet. E) and G) proteins are matched to known structural models in the Protein Data Base (PDB) and modeled by threading techniques. ($G_{1-2}$) Molecular simulation techniques, such as NAMD and CHARM, model the proteins in dynamic conditions. (H) Analysis of protein models and physico-chemical properties can pinpoint key amino acid sequence differences and structural variations that are predicted to be involved in providing the proteins with their acid stable properties.

## 3. Human Resource Development

### 3.1 Teaching and Seminars

Investigators at the participating institutions cross-teach courses in bioinformatics, genome biology and protein simulation at the undergraduate and graduate level. Seminars and special events are shared via video conferencing.

### 3.2 Shared Doctoral Students

Whereas shared teaching and seminars have had a marked impact on promoting collaborative learning and investigation between the participating institutions, the main catalytic element driving the cohesion of the VIIB project has proved to be the sharing of doctoral students. Investigators from participating institutions are co-tutors of doctoral students and students' thesis committees contain personal from the different institutions.

Advances and exams of doctoral students are carried out via video conferencing, overcoming problems of geographical separation. It is anticipated that the VIIB will play a substantial role in providing mechanisms for cross-disciplinary training and collaboration and educational programs; all toward a future of e-Science.

## 4. Outreach

The VIIB considers that outreach programs to schools and to the public are an important part of its mission. It has conducted several bioinformatics workshops for high school teachers and personnel from the VIIB have presented training sessions in bioinformatics and genome biology at high schools. Since VIIB is a Latin American initiative, some components of the teaching and training, especially those carried out in middle school and junior high school, are carried out in Spanish. However, research activities and doctoral student presentations are in English. The reasoning is that we believe that Latin American scientific and engineering training should result in individuals who are bilingual. Also, since most of the relevant web pages are in English it becomes import to be bilingual even at the school level. Interestingly, VIIB activities have proved to be an important driving force for students to learn English.

A new outreach program called "Adopt-a-Gene" is being developed in which the VIIB supplies schools and colleges with DNA sequences encoding individual genes, multigene segments or even whole genomes, depending on the interest and level of sophistication of the schools. The idea is that the students annotate the gene(s) and search for information that describes the biology of the gene(s). The VIIB supplies detailed tutorials to aid the students and also provides tutors online who help answer questions. The idea is to present learning opportunities in bioinformatics and genome biology to young people who are typically very proficient in internet use. Schools that are remote or have underprivileged students will be especially targeted by the initiative.

## 5. Cyberinfrastructure Available at the VIIB

The VIIB connects three computer clusters that provide a combined total of about 300 cpus for parallel processing. This will soon be upgraded to about 600 cpus. There are also facilities for the simultaneous 3-D viewing of molecules for 20 people. The VIIB is connected to the Chilean internet 2 fiberoptic backbone network supported by REUNA that, in turn, is connected to various portals in Latin America, North America via RedCLARA and to Europe via ALICE (Figure 4).

## 6. E-Science Challenges

### 6.1 Interoperabililty and Connectivity Concerns

The VIIB operates in several computational environments including UNIX/Linux/Windows/OS X. This has proved a challenge for interoperability both within and between the constituent groups of the VIIB. Condor has provided an acceptable solution as a middleware. The adoption of Condor is also important if the VIIB becomes connected with other large scale e-science initiatives such as EELA that frequently use Globus as middleware. Unlike Condor, Globus is not compatible at the moment with the Microsoft Windows cluster installed at the CBGB. Web services are also being developed to overcome interoperability problems.

High speed connectivity within Chile via REUNA and within Latin America via

RedCLARA is more expensive than counterparts in North America and Europe. The continued efforts of the VIIB will require stable funding to maintain and enhance the connectivity that has been established.

### 6.2 High Performance Computing Limitations

The combined computational power of the VIIB is already strained in terms of use, leading to problems of scalability and scheduling. No new projects can be contemplated until the computational capacity is upgraded via addition of new clusters or by accessing and sharing existing large-scale multinational grid computing networks. Both options are being pursued. The computational power of the VIIB will soon be upgraded to about 600 cpus, doubling its present capacity and the VIIB has applied for entry into the EELA 2 network that will potentially expand its computational capabilities.

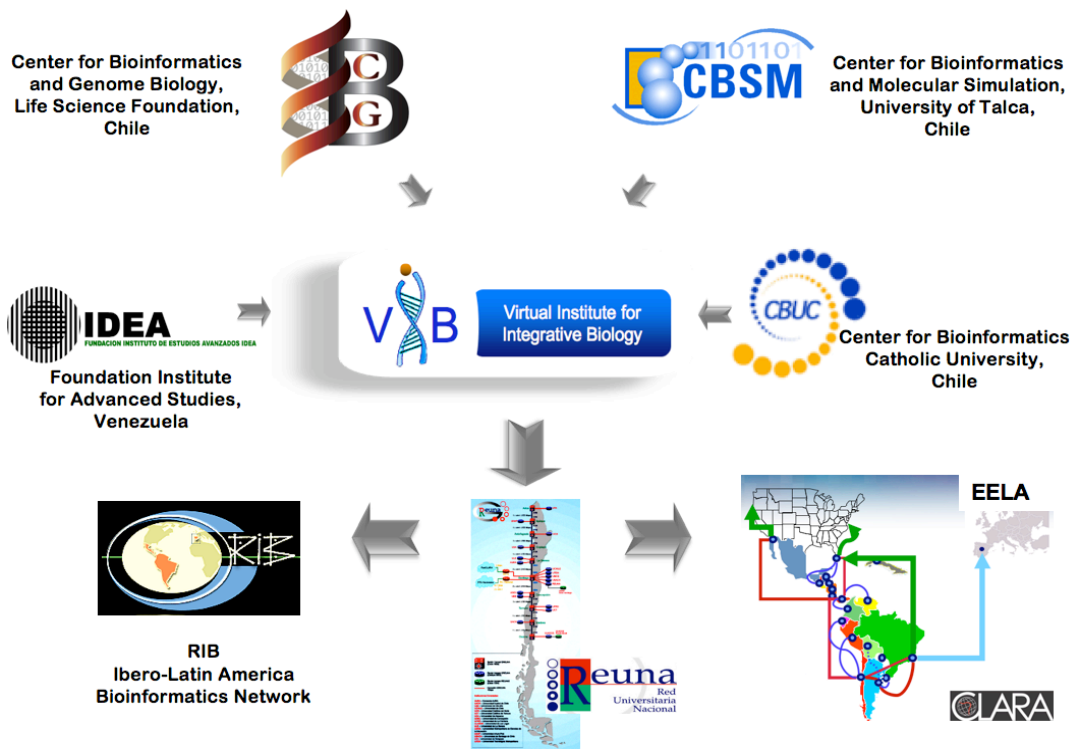

**Figure 4. Network connections of the VIIB.** Connectivity is shown between the four cornerstone institutes via the VIIB and the VIIB also connects to other Latin America and North American institutes via REUNA and RedClara. It is hoped that the VIIB will join the EELA 2 European network.

### 6.3 Development of Customized Computational Frameworks and Flexible Workflows

Each project addresses different scientific questions and uses a different but overlapping suite of bioinformatics algorithms and interaction and visualization tools. As such each workflow has had to be customized to fulfil the specific demands of the project. The TAVERNA workbench [8] has proved a useful and flexible paradigm for distributed workflow using the SOAP server [9], but other options are being explored such as PEGASYS [10] together with enhancements to SOAP to process larger databases. A key problem has been to efficiently exploit shared resources without causing impediments to the user who has little interest in the underlying information technology (IT). We are still a long way from providing an optimal user-friendly, intuitive workspace for the biologist.

**6.4 Sharing, Storage and Dissemination of Information**

The essence of an e-science colaboratory is the sharing of primary data, techniques and results. The VIIB considers Open Source to be mission-critical. Databases and access tools are mirrored at the participating institutions and publications are also being made available within the VIIB. Discussion continues regarding the free access of courses because the constituent institutes are connected to private universities who are thus stakeholders in the educational resources offered by the VIIB. However, progress is being made in that a doctoral course in bioinformatics will be soon be made freely accessible via the web using the MIT model.

**6.5 Improving Visibility**

It is hoped that the combination of good science and open access to data and publications will expand the visibility of the VIIB in the world scientific community. The benefits to the VIIB are foreseen to be: (i) improving the potential to incorporate new partners, (ii) promoting access to additional funding opportunities and research initiatives and (iii) improving the potential for students and post-doctoral associates to be placed in world class laboratories.

## 7. Next Steps

- Improve intra- and inter-operability via the development of web services.
- Improve computational power by connecting to multinational grids.
- Expand the outreach program to schools and colleges via the Adopt-a-Gene program. Mobile phone devises are being considered to access remote areas in Latin America.
- Develop new initiatives in nanoinformatics in the area of nanobiotechnology.
- Incorporate new academic and industrial partners both in Latin America and in the rest of the world.

## 8. Summary and Conclusions

The VIIB has proved an effective way for small teams to transcend the critical mass problem, to overcome geographic limitations, to harness the power of large scale, collaborative science and to improve the visibility of Latin American science. It is hoped that a combination of good science and open access will continue to expand this visibility providing, in the long term, additional research opportunities and promoting the recruitment of additional scientific partners throughout Latin America and the world. As such, the VIIB may provide a useful paradigm for encouraging further e-science initiatives in Latin America and other emerging regions.

## Acknowledgments

Support was provided by a Microsoft Sponsored Research Award, Fondecyt 1050063 the Millennium Institute for Fundamental and Applied Biology, the Bicentenary Program for Science and Technology ACT/24 and Fundación Chilena para Biología Celular.